\documentclass[12pt,titlepage]{article}
\usepackage{amsmath}
\usepackage{amssymb}
\usepackage[dvips]{graphicx}
\usepackage{amsfonts}
\usepackage[portuguese]{babel}

\begin{document}

\title{Potenciais delta revisitados\\
via transformada de Fourier\thanks{To appear in Revista Brasileira de Ensino de F\'{\i}sica}\bigskip \\
{\small \ (Delta potentials revisited via Fourier transform)}}
\author{A.S. de Castro\thanks{%
E-mail: castro@pq.cnpq.br} \\
\\
Departamento de F\'{\i}sica e Qu\'{\i}mica, \\
Universidade Estadual Paulista \textquotedblleft J\'{u}lio de Mesquita
Filho\textquotedblright, \\
Guaratinguet\'{a}, SP, Brasil}
\date{}
\maketitle

\begin{abstract}
O problema de estados ligados em potenciais delta \'{e} revisitado usando
uma abordagem baseada na transformada de Fourier. O problema de um simples
potencial delta resume-se \`{a} resolu\c{c}\~{a}o de uma equa\c{c}\~{a}o alg%
\'{e}brica de primeiro grau para a transformada de Fourier da autofun\c{c}%
\~{a}o e o problema para mais que uma fun\c{c}\~{a}o delta tamb\'{e}m
revela-se uma quest\~{a}o simples. Diferentemente de m\'{e}todos diretos,
nenhum conhecimento acerca da descontinuidade de salto da derivada primeira
da autofun\c{c}\~{a}o \'{e} necess\'{a}rio para determinar a solu\c{c}\~{a}o
do problema.\newline
\newline
\noindent \textbf{Palavras-chave:} delta de Dirac, estado ligado,
transformada de Fourier.\newline
\newline
\newline

{\small \noindent The problem of bound states in delta potentials is
revisited by means of Fourier transform approach. The problem in a simple
delta potential sums up to solve an algebraic equation of degree one for the
Fourier transform of the eigenfunction and the problem for more than one
delta function also reveals itself to be a simple matter. Quite differently
from direct methods, no knowledge about the jump discontinuity of the first
derivative of the eigenfunction is required to determine the solution of the
problem. \newline
\newline
}

\end{abstract}

A an\'{a}lise da equa\c{c}\~{a}o de Schr\"{o}dinger com um potencial constitu%
\'{\i}do de uma soma de duas fun\c{c}\~{o}es delta de Dirac \ ocupa as p\'{a}%
ginas de muitos livros-texto \cite{gas}-\cite{gri}. Os poss\'{\i}veis
estados ligados s\~{a}o encontrados pela localiza\c{c}\~{a}o dos polos
complexos da amplitude de espalhamento ou por meio de uma solu\c{c}\~{a}o
direta da equa\c{c}\~{a}o de Schr\"{o}dinger com fulcro na descontinuidade
de salto da derivada primeira da autofun\c{c}\~{a}o, mais a continuidade da
autofun\c{c}\~{a}o e seu bom comportamento assint\'{o}tico. Em um trabalho
recente \cite{asc} tal problema foi examinado com o m\'{e}todo da
transformada de Laplace. Resultou que a solu\c{c}\~{a}o do problema de
estados ligados n\~{a}o requer qualquer conhecimento sobre a descontinuidade
da derivada primeira da autofun\c{c}\~{a}o. Na esteira da Ref. \cite{asc}, o
problema de estados ligados em potenciais delta de Dirac \'{e} agora
revisitado usando a abordagem via transformada de Fourier. O problema com um
potencial constitu\'{\i}do de uma \'{u}nica fun\c{c}\~{a}o delta resume-se a
resolver uma equa\c{c}\~{a}o alg\'{e}brica de primeira ordem para a
transformada de Fourier da autofun\c{c}\~{a}o, e o problema para mais que
uma fun\c{c}\~{a}o delta tamb\'{e}m revela-se desembara\c{c}ado. Sucede que,
tal como na an\'{a}lise via transformada de Laplace, o conhecimento da
descontinuidade da derivada primeira da autofun\c{c}\~{a}o \'{e}
irrelevante. Para dizer a verdade, a transformada de Fourier tem sido usada
para abordar os estados ligados do oscilador harm\^{o}nico qu\^{a}ntico de
uma maneira simples e elegante \cite{mu}-\cite{pr}. Mais recentemente,
baseado em um \textit{ansatz} para o comportamento assint\'{o}tico da autofun%
\c{c}\~{a}o, Palm and Raff \cite{pr} desenvolveram um m\'{e}todo para lidar
com uma classe ampla de potenciais via transformada de Fourier. Entretanto,
os potenciais contemplados pelo m\'{e}todo constante na Ref. \cite{pr} s\~{a}%
o aqueles constitu\'{\i}dos de uma soma de termos da forma $x^{\gamma }$,
com $-2\leq \gamma \leq 2$, e assim sendo o procedimento de Palm and Raff
\'{e} inepto para resolver problemas envolvendo deltas de Dirac.

Come\c{c}aremos nosso exame com um \'{u}nico delta de Dirac localizado na
origem. A equa\c{c}\~{a}o de Schr\"{o}dinger independente do tempo para o
potencial delta \'{e} dada por
\begin{equation}
\left[ -\frac{\hbar ^{2}}{2m}\frac{d^{2}}{dx^{2}}-\alpha \delta \left(
x\right) \right] \phi \left( x\right) \,=E\phi \left( x\right) ,  \label{eq1}
\end{equation}%
onde $\alpha $ \'{e} um par\^{a}metro real. Usando as defini\c{c}\~{o}es%
\begin{equation}
a=\frac{2m\alpha }{\hbar ^{2}},\qquad b=\sqrt{\frac{2m|E|}{\hbar ^{2}}},
\label{ab}
\end{equation}%
a equa\c{c}\~{a}o para estados ligados ($E=-|E|$) pode ser escrita como%
\begin{equation}
\frac{d^{2}\phi \left( x\right) }{dx^{2}}+a\delta \left( x\right) \phi
\left( x\right) -b^{2}\phi \left( x\right) =0.  \label{eq2}
\end{equation}%
Em raz\~{a}o da paridade par da fun\c{c}\~{a}o delta de Dirac, i.e. $\delta
\left( -x\right) =\delta \left( x\right) $, a autofun\c{c}\~{a}o pode ser
escolhida para ser par ou \'{\i}mpar. Definindo $\Phi \left( k\right) $ como
a transformada de Fourier de $\phi \left( x\right) $,
\begin{equation}
\Phi \left( k\right) =\mathcal{F}\left\{ \phi \left( x\right) \right\} =%
\frac{1}{\sqrt{2\pi }}\int_{-\infty }^{+\infty }dx\,e^{ikx}\phi \left(
x\right) ,
\end{equation}%
a transformada de Fourier inversa \'{e} dada por%
\begin{equation}
\phi \left( x\right) =\mathcal{F}^{-1}\left\{ \Phi \left( k\right) \right\}
\frac{1}{\sqrt{2\pi }}\int_{-\infty }^{+\infty }dk\,e^{-ikx}\Phi \left(
k\right) .
\end{equation}%
Pode ser mostrado que $\Phi ^{\ast }\left( k\right) =\Phi \left( -k\right) $
se $\phi \left( x\right) $ for real e $\Phi ^{\ast }\left( k\right) =-\Phi
\left( -k\right) $ se $\phi \left( x\right) $ for imagin\'{a}rio \cite{but}.
Com estes resultados, e porque qualquer fun\c{c}\~{a}o pode ser expressa
como uma combina\c{c}\~{a}o linear de suas partes real e imagin\'{a}ria,
pode-se perceber que a paridade de $\Phi \left( k\right) $ sob a troca de $k$
por $-k$ \'{e} a mesma que essa de $\phi \left( x\right) $ sob a troca de $x$
por $-x$. Assumindo que $\phi \left( x\right) $ e sua derivada se anulam
quando $|x|\rightarrow \infty $, pode-se escrever%
\begin{equation}
\mathcal{F}\left\{ \frac{d^{2}\phi \left( x\right) }{dx^{2}}\right\} =-k^{2}%
\mathcal{F}\left\{ \phi \left( x\right) \right\} ,
\end{equation}%
de modo que a transformada de Fourier de (\ref{eq2}) nos conduz a uma equa%
\c{c}\~{a}o alg\'{e}brica para $\Phi \left( k\right) $ cuja solu\c{c}\~{a}o
\'{e}
\begin{equation}
\Phi \left( k\right) =a\,\frac{\phi \left( 0\right) }{\sqrt{2\pi }}\,\frac{1%
}{k^{2}+b^{2}}.
\end{equation}%
Haja vista que $\Phi \left( k\right) $ \'{e} uma fun\c{c}\~{a}o par pode-se
concluir que $\phi \left( x\right) $ \'{e} tamb\'{e}m uma fun\c{c}\~{a}o
par. Li\c{c}\~{o}es de Mec\^{a}nica Qu\^{a}ntica (\textit{...if the
eigenfunctions are arranged in the order of increasing eigenvalues of the
energy, these functions are alternately even and odd, the eigenfunction of
the ground state is always even} \cite{mes}) permite-nos especular, sem
qualquer maquin\'{a}rio matem\'{a}tico adicional, que a solu\c{c}\~{a}o de
nosso problema, se \'{e} que ela existe, \'{e} unica. Pode ser mostrado que
\cite{but}
\begin{equation}
\mathcal{F}\left\{ e^{-\Gamma |x|}\right\} =\sqrt{\frac{2}{\pi }}\,\frac{%
\Gamma }{k^{2}+\Gamma ^{2}},\qquad \Gamma >0.  \label{lor1}
\end{equation}%
Usando este fato pode-se escrever%
\begin{equation}
\phi \left( x\right) =\frac{a}{2b}\,\phi \left( 0\right) e^{-b|x|},
\label{psi}
\end{equation}%
que \'{e} v\'{a}lido somente se $a/\left( 2b\right) =1$. Obviamente $\phi
\left( x\right) $ \'{e} uma solu\c{c}\~{a}o aceit\'{a}vel somente para $a>0$%
. Conforme j\'{a} especulado pelo uso de argumentos de simetria, existe uma
e somente uma solu\c{c}\~{a}o de estado ligado. Em termos das vari\'{a}veis
originais, esta solu\c{c}\~{a}o \'{u}nica pode ser expressa como%
\begin{equation}
E=-\frac{m\alpha ^{2}}{2\hbar ^{2}},\qquad \phi \left( x\right) =\,\phi
\left( 0\right) e^{-\frac{m\alpha }{\hbar ^{2}}|x|},\qquad \alpha >0.
\end{equation}

Agora consideraremos a equa\c{c}\~{a}o de Schr\"{o}dinger com duas fun\c{c}%
\~{o}es delta separadas pela dist\^{a}ncia $2L$:

\begin{equation}
\frac{d^{2}\phi \left( x\right) }{dx^{2}}+a\left[ \delta \left( x+L\right)
+\delta \left( x-L\right) \right] \phi \left( x\right) -b^{2}\phi \left(
x\right) =0,
\end{equation}%
com $a$ e $b$ definidos como antes. Neste caso, $\Phi \left( k\right) $
torna-se
\begin{equation}
\Phi \left( k\right) =\frac{a}{\sqrt{2\pi }}\left[ \phi \left( -L\right)
e^{-ikL}\frac{1}{k^{2}+b^{2}}+\phi \left( +L\right) e^{+ikL}\frac{1}{%
k^{2}+b^{2}}\right] .
\end{equation}%
\'{E} instrutivo observar que $\Phi \left( k\right) $ \'{e} uma fun\c{c}\~{a}%
o par ou \'{\i}mpar sob a troca de $k$ por $-k$ consoante $\phi \left(
x\right) $ seja par ou \'{\i}mpar, respectivamente, de forma que o problema
possivelmente admite autofun\c{c}\~{o}es pares ou \'{\i}mpares. Usando a
propriedade de deslocamento das transformadas de Fourier \cite{but}
\begin{equation}
\mathcal{F}\left\{ \phi \left( x\pm L\right) \right\} =e^{\mp ikL}\mathcal{F}%
\left\{ \phi \left( x\right) \right\} ,
\end{equation}%
se obt\'{e}m%
\begin{equation}
\Phi \left( k\right) =\frac{a}{2b}\left[ \phi \left( -L\right) \mathcal{F}%
\left\{ e^{-b|x+L|}\right\} +\phi \left( +L\right) \mathcal{F}\left\{
e^{-b|x-L|}\right\} \right] ,
\end{equation}%
e usando (\ref{lor1}) outra vez, $\phi \left( x\right) $ \'{e} reconstru%
\'{\i}da como%
\begin{equation}
\phi \left( x\right) =\frac{a}{2b}\left[ \phi \left( -L\right)
e^{-b|x+L|}+\phi \left( +L\right) e^{-b|x-L|}\right] .
\end{equation}%
A continuidade de $\phi \left( x\right) $ em $x=L$ implica em%
\begin{equation}
e^{-2bL}=\left( \frac{2b}{a}-1\right) \frac{\phi \left( +L\right) }{\phi
\left( -L\right) }.  \label{qua}
\end{equation}%
As poss\'{\i}veis solu\c{c}\~{o}es desta condi\c{c}\~{a}o de quantiza\c{c}%
\~{a}o podem ser visualizadas graficamente por meio de esbo\c{c}os de seus
membros direito e esquerdo como fun\c{c}\~{a}o de $bL$. As abscissas das
interse\c{c}\~{o}es fornecem as solu\c{c}\~{o}es. Pode-se inferir que n\~{a}%
o existe nenhuma solu\c{c}\~{a}o para $a<0$. Quanto a $a>0$, sempre existe
uma solu\c{c}\~{a}o com $E<-m\alpha ^{2}/\left( 2\hbar ^{2}\right) $. A exist%
\^{e}ncia de uma solu\c{c}\~{a}o (\'{\i}mpar) adicional, em mais alta
energia, ocorre somente se $aL>1$ \cite{tam}. Este limiar ocorre porque as
curvas $e^{-2bL}$ e $1-2b/a$ se osculam em $bL=0$ quando $aL=1$ e se
interceptam em algum ponto com abscissa $bL>0$ se e somente se $aL>1$.

A continuidade de $\phi (x)$ e a magnitude da descontinuidade de salto de
sua derivada primeira s\~{a}o ingredientes essenciais para resolver a equa%
\c{c}\~{a}o de Schr\"{o}dinger com potenciais delta de Dirac pela for\c{c}a
bruta. A abordagem via transformada de Fourier, contudo, somente requer que $%
\phi (x)$ e sua derivada primeira se anulem quando $|x|\rightarrow \infty $.
O ajuste trivial, $\underset{x\rightarrow 0}{\lim }\;\phi (x)=\phi (0)$,
\'{e} suficiente para determinar a solu\c{c}\~{a}o no caso de uma simples fun%
\c{c}\~{a}o delta, e a continuidade de $\phi (x)$ em $x=L$ \'{e} necess\'{a}%
ria no caso de um delta duplo.

A generaliza\c{c}\~{a}o para um potencial peri\'{o}dico formado por uma sequ%
\^{e}ncia de fun\c{c}\~{o}es delta igualmente espa\c{c}adas \'{e} deixada
para os leitores.

\vspace{10cm}

\noindent{\textbf{Agradecimentos}}

O autor \'{e} grato ao CNPq pelo apoio financeiro.

\end{document}